\documentclass{aa}  

\usepackage{graphicx}
\usepackage[normalem]{ulem}
\usepackage{amsmath}
\usepackage{color}
\usepackage{txfonts}
\usepackage[]{hyperref}

\usepackage{natbib,twoopt}

\newcommand{\angstrom}{\mbox{\normalfont\AA}}

\begin{document} 
   \title{Possible regular phenomena in EXO 2030+375}
   

   \author{Eva Laplace
		   \inst{1, 2}
		   \and
		   Tatehiro Mihara
		   \inst{2}
		   \and 
		   Yuki Moritani
		   \inst{3}
		   \and
		   Motoki Nakajima
		   \inst{4}
		   \and
		   Toshihiro Takagi
		   \inst{2}
		   \and
		   Kazuo Makishima
		   \inst{2}
		   \and
		   Andrea Santangelo
		   \inst{1}
          }

   \institute{Institut für Astronomie und Astrophysik, University of Tübingen, Sand 1, 72076 Tübingen, Germany\\
              \email{laplace@astro.uni-tuebingen.de}\\
          \and
             MAXI team, RIKEN, 2-1 Hirosawa, Wako, Saitama 351-0198, Japan\\
             \email{tmihara@riken.jp}\\
          \and
	         Kavli Institute for the Physics and Mathematics of the Universe (WPI),
	         The University of Tokyo Institutes for Advanced Study, The University of Tokyo, Kashiwa, Chiba 277-8583, Japan\\
          \and
	         School of Dentistry at Matsudo, Nihon University, 2-870-1, Sakaecho-nishi, Matsudo, Chiba 271-8587, Japan}

	\date{Received 22 July 2016 / Accepted 23 September 2016}
	
 \abstract{ 
In the last 10 years, since its last giant outburst in 2006, regular X-ray outbursts (type I) were detected every periastron passage in the Be X-ray binary EXO 2030+375. Recently, however, it was reported that the source started to show a peculiar behavior: its X-ray flux decreased significantly and type I outbursts were missed in several cases. At the same time, the spin frequency of the neutron star, which had been increasing steadily since the end of the 2006 giant outburst, reached a plateau. Very recent observations indicate that the source is now starting to spin down.\\
These observed phenomena have a striking similarity with those which took place 20 years ago, just before the source displayed a sudden orbital phase shift of the outburst peak (1995). This historical event occurred at the time exactly between the two giant outbursts (1985 and 2006). These phenomena suggest the system to have an underlying periodicity of 10.5 years between orbital phase shifts and/or giant outbursts. The suggested periodicity may reflect some long-period dynamics in the circumstellar disk of the Be star, due, e.g., to the Kozai-Lidov effect. A model generating such a periodic change of the Be disk, namely Kozai-Lidov oscillations in the Be disk, is discussed. If this behavior is really periodical, another phase shift of the X-ray outburst peak is predicted to occur around 2016 December.}
 

   \keywords{	accretion, accretion disks -- stars: neutron, emission-line, Be -- stars: individual: EXO 2030+375 --
                X-rays: binaries
               }

   \maketitle
%

\section{Introduction}
Be X-ray binaries (hereafter BeXRBs) are binary star systems composed of a neutron star (typically with a strong magnetic field) and a (O9-B2)-type star from which emission lines have been observed, known as a Be star. From optical and infrared observations, these stars are known to harbor large circum-stellar disks, which also provide a satisfactory explanation of the X-ray radiation observed from these systems \citep{reig_be/x-ray_2011}. Most BeXRBs exhibit a characteristic transient X-ray radiation, so-called outbursts, during which the X-ray luminosity increases until it reaches a maximum and fades away again, on a timescale from a week to several months. Two main types of X-ray outbursts are commonly distinguished: (i) type I, also known as normal outbursts, with a typical luminosity of $L_\mathrm{X}\sim 10^{36}-10^{37}~\mathrm{\mbox{erg}}\mathrm{\mbox{ s}}^{-1}$, which typically occur periodically at the periastron passage time, hence with a period equal to the orbital period; (ii) type II or giant outbursts, which are significantly more luminous with $L_\mathrm{X}\ga10^{37}~\mathrm{\mbox{erg}}\mathrm{\mbox{ s}}^{-1}$ and can last for several orbital periods. Giant outbursts are generally assumed to occur randomly \citep{stella_intermittent_1986, okazaki_natural_2001}.\\
The origin of the X-ray radiation from type I outbursts is usually explained by the interaction between the Be-star disk and the neutron star. When the neutron star approaches the Be star and its disk closely enough, accretion can occur, resulting in the emission of X-ray radiation. Similarly, type II outbursts are considered to arise through accretion from the disk, but it is still unclear what causes them to be different from type I outbursts, including in parts accretion at times differing from the periastron passage. Latest theories and simulations suggest that an increased eccentricity or warping of the Be disk, which is misaligned with the orbital plane, cause the type II outbursts \citep{martin_giant_2014}. \citet{okazaki_origin_2013} found that enormous amounts of matter can be transferred through Bondi-Hoyle Lyttleton accretion from the warped Be disk in misaligned systems.\\ 
\\
The X-ray source \object{EXO 2030+375} is a BeXRB, composed of a magnetized neutron star and a B0 Ve companion \citep{coe_optical/ir_1988}. It was discovered on 1985 May 17 during a giant outburst \citep{parmar_transient_1989}. The source is characterized by a spin period of 42 s, an orbital period of 46.021 days, an eccentricity of 0.4190 \citep{wilson_outbursts_2008} and by being the Be X-ray binary with the largest number of observed type I outbursts (over 150), occurring almost every periastron passage.\\
Owing to its nearly continuous activity, EXO 2030+375 has been studied extensively. Ten years after the discovery giant outburst, the source experienced an orbital phase jump (hereafter the OPJ95) of the type I outbursts around MJD 50000 in 1995 \citep{reig_timing_1998, wilson_decade_2002}. The outburst peak of the next detected outburst was shifted by about 8-9 days earlier than the preceding outbursts. Ten years after this event, a second giant outburst was observed in 2006 \citep{corbet_outburst_2006, klochkov_integral_2007, baykal_recent_2008, wilson_outbursts_2008}. The type I outbursts after the giant outburst were shifted by about 8-9 days later than the previous type I outbursts (hereafter the OPJ06).\\
Since these events, the source has been continuously monitored by the Monitor of All-sky X-ray Image (MAXI), a Japanese instrument onboard the International Space Station, the Burst Alert Telescope (BAT) on the Swift satellite, and the Gamma-ray Burst Monitor (GBM) on the Fermi satellite. As a result, the X-ray flux and spin period changes of the source were determined.\\
After the last giant outburst, a type I outburst was detected at every periastron passage, at least until 2012 July.
Recently, however, the source faded in X-rays, and some outbursts were missed. Simultaneously, the spin period, which had been increasing since the 2006 giant outburst, reached a plateau \citep{atel_fading}. Using Fermi/GBM data, we report that the source is now starting to spin down.\\ 
A single measurement of the H$\alpha$ line was taken on 2014 December 7 (Iaian Steele, private communication). Its equivalent width is of $11\pm2\mbox{ \angstrom}$, similar to the profiles before the OPJ95 (around $15\mbox{ \angstrom}$).\\

These phenomena resemble those before the OPJ95. They suggest that giant outbursts and orbital phase shifts are linked and occur every 20 years.\\
In this paper, we investigate the recent phenomena and the possible periodicity by analyzing the X-ray monitoring data, the spin period, and the H$\alpha$ equivalent width in the context of the entire historical behavior of EXO 2030+375.
\section{Observations}
\subsection{MAXI/GSC}
The Monitor of All-sky X-ray Image (MAXI) is an instrument onboard the International Space Station (ISS) that has been continuously monitoring X-ray sources since the beginning of its operations in 2009 August \citep{matsuoka_maxi_2009}. Its main instrument, the Gas Slit Camera (GSC), is sensitive to an energy range of 2-30 keV and has a time resolution of 0.1 ms \citep{mihara_gas_2011}. Every 92 minutes, the ISS completes an orbit around the Earth, which enables MAXI to observe the entire X-ray sky, except regions close to the Sun. As a result, the MAXI/GSC data are useful to study the long-term variations of X-ray sources. We used the publicly available light-curves of the official MAXI website\footnote{\url{http://maxi.riken.jp/top/}} for this study.
\subsection{Swift/BAT} 
Swift was launched in 2004 and designed specifically for the search
of Gamma-ray bursts. Its monitoring instrument, the Burt Alert Telescope (BAT), is searching for transient
events in the 15-50 keV band from 80\% of the sky \citep{barthelmy_burst_2005}. Light curves for monitored sources are publicly available on the official website\footnote{\url{http://swift.gsfc.nasa.gov/results/transients/}}, the Hard X-rays Transient Monitor \citep{krimm_swift/bat_2013}.
\subsection{Fermi/GBM}
The Fermi Gamma-Ray Space Telescope was launched in 2008 and is dedicated to the study of gamma-ray sources. Its Gamma-ray Burst Monitor (GBM) instrument has an energy range of 20 MeV to about 300 GeV \citep{meegan_fermi_2009}.  For this study, the public pulsar data\footnote{\url{http://gammaray.nsstc.nasa.gov/gbm/science/pulsars.html}} were used.
\subsection{RXTE/ASM}
The Rossi X-ray Timing Explorer was launched in December 1995 and was operative until January 2012. Its All Sky Monitor instrument had a total collecting area of $90 \mbox{\textrm{ cm}}^2$. It had a monitoring capability of 80\% of the sky per minute and was operative in the 2-10 keV band \citep{levine_first_1996}. Public archival data from this instrument\footnote{\url{http://heasarc.gsfc.nasa.gov/docs/xte/ASM/sources.html}} were used for the study.
\section{Data analysis}
\begin{figure*}
\centering
\includegraphics[width=23cm, angle=90]{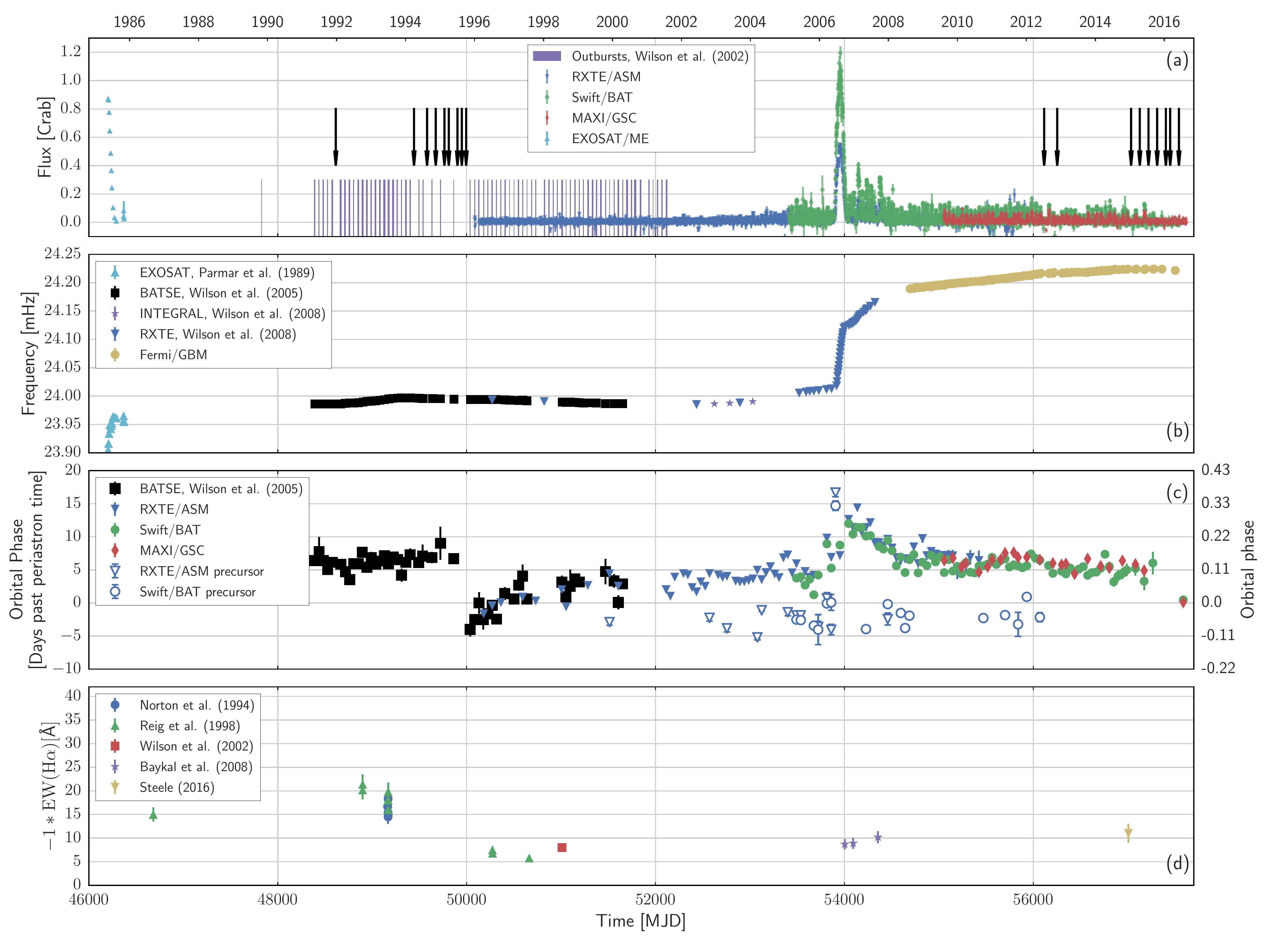}
\caption{Overview of the long-term evolution of EXO 2030+375. (a) Long-term history of the X-ray light curve rescaled to the Crab flux; black arrows indicate missed type I outbursts. Outbursts listed in \citet{wilson_decade_2002} are indicated as violet rectangles. The width of the rectangle corresponds to the outburst timescale. (b) Long-term history of the spin period from \citet{wilson_discovery_2005,wilson_outbursts_2008} and Fermi/GBM data. The error bars are smaller than the symbols. (c) Peak outburst time with respect to the periastron passage time and peak precursor peak time with respect to periastron passage time determined with the method described in the text. (d) Long-term history of the H$\alpha$ line equivalent width from various authors. Detailed information is given in the text.}
\label{fig:total_plot}
\end{figure*}
\subsection{Overview}
Recent monitoring data of several instruments are combined with previously published X-ray and optical data to understand the recent behavior of EXO 2030+375 in the context of its history.\\
An overview of the entire behavior of the source can be found in Fig. \ref{fig:total_plot}, where the light curve and the evolution of the spin frequency, the timing of the peak, and the equivalent width of the H$\alpha$ line are shown.
\subsection{Light curve}
In the light curve we show in Fig. \ref{fig:total_plot} panel (a), EXOSAT/ME data from 
\citet{parmar_transient_1989} and monitoring data from RXTE/ASM, Swift/BAT and MAXI/GSC are plotted. For better comparison, the flux of each instrument was rescaled to the Crab flux with a value of $1 \textrm{ Crab} = 0.22 \textrm{ counts s}^{-1}\textrm{ cm}^{-2}$ \citep{krimm_swift/bat_2013}, $1 \textrm{ Crab} = 3.6 \textrm{ counts s}^{-1}\textrm{ cm}^{-2}$ \citep{matsuoka_maxi_2009} and $1 \textrm{ Crab} = 75.5 \textrm{ counts s}^{-1}\textrm{ cm}^{-2}$ \citep{levine_first_1996} for Swift/BAT (15-50 keV), MAXI (2-20 keV) and RXTE/ASM (1-12 keV), respectively. Black arrows indicate the times at which no significant X-ray emission was found during the predicted type I outbursts \citep[orbital solution of][]{wilson_outbursts_2008} from the BATSE outbursts listed in \citet{wilson_decade_2002} (indicated with violet rectangles) and from recent monitoring data.\\
It is apparent that outbursts are missed more and more frequently for the two periods in which there were missing outbursts (in 1995 and 2015). 
\subsection{Spin frequency}
The evolution of the spin frequency is shown in Fig. \ref{fig:total_plot} (b). Data from \citet{wilson_outbursts_2008} are shown as well as recent Fermi/GBM data. During the two giant outbursts, a dramatic spin-up of the neutron star was observed (MJD~46200 and MJD~54000). The subsequent transition to a slower spin up followed by a constant spin is remarkably similar both around MJD 49300, before the pulsar started spinning down and for the most recent measurements.\\
The latest data points strongly indicate that the source is now starting to spin down (see also Fig. \ref{fig:spin_down}).
\subsection{H$\alpha$ line equivalent width}
In BeXRBs, crucial information about the Be disk size and state can be obtained from optical observations. The most prominent spectral emission line in the optical band is the hydrogen $\alpha$-line, and its equivalent width is a good indication of the Be disk size \citep[see, e.g.][]{reig_be/x-ray_2011}. In Fig. \ref{fig:total_plot} (d), we combine all published measurements of the H$\alpha$ line equivalent width \citep{norton_multiwavelength_1994,reig_long-term_1998,wilson_decade_2002,baykal_recent_2008}. At present as before the OPJ95, an increase of the equivalent width can be seen. A dramatic drop of the equivalent width occurred after the OPJ95 (MJD~50500). Because of lacking observations, it remains unfortunately unclear how it evolved earlier. However, infrared photometry measurements indicate a drop in magnitude at least 100 days before the OPJ95 \citep{reig_long-term_1998,wilson_decade_2002}.\\
\subsection{Timing of the outburst peak}
\label{section:peakTiming}
\begin{figure}
\resizebox{\hsize}{!}{\includegraphics{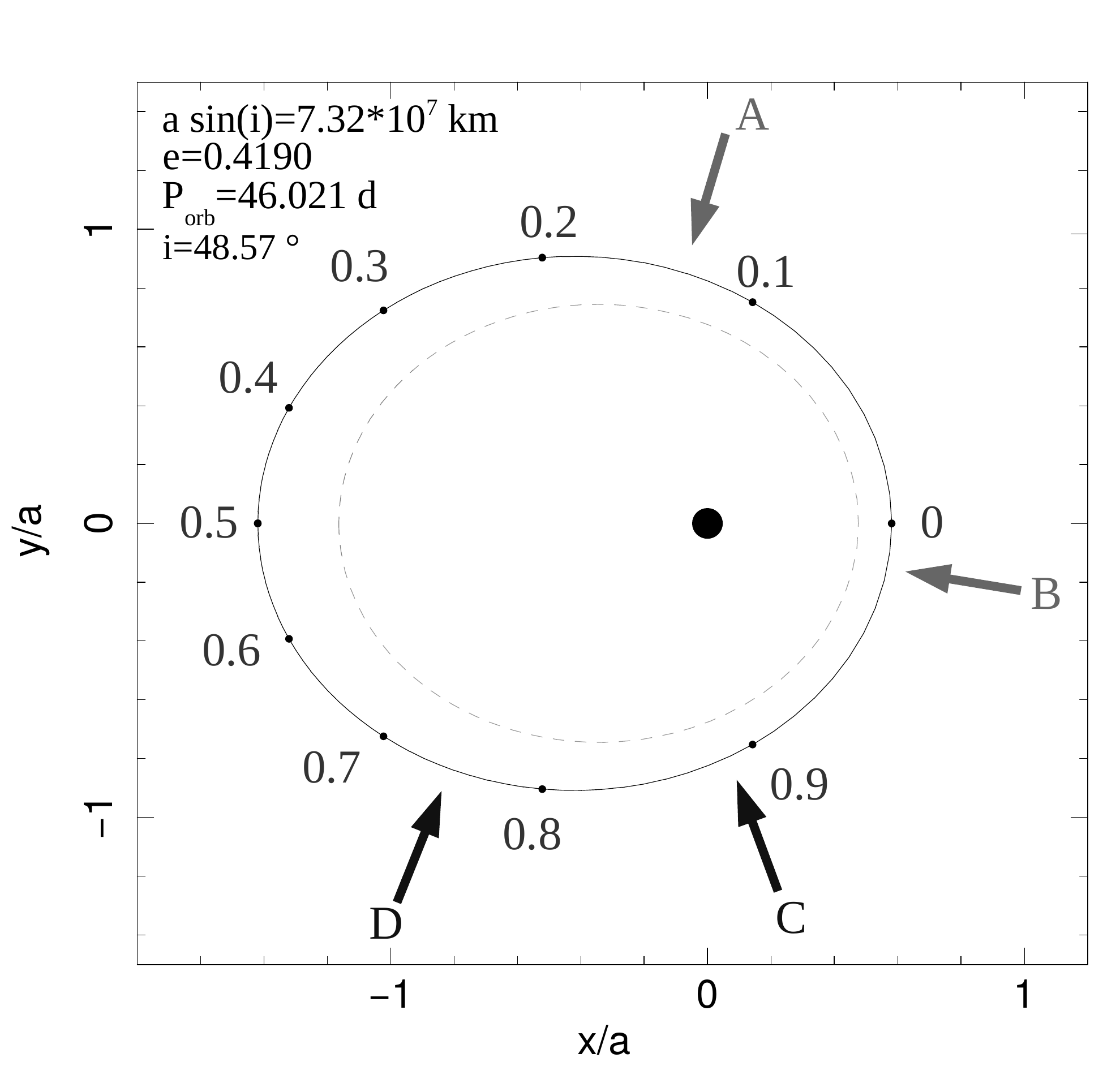}}
\caption{Calculated orbit using the parameters from \citet{wilson_outbursts_2008}, displayed in the top right corner. The central black dot, the solid line, and the dashed line represent the Be star, the neutron star orbit, and the L1 point trajectory, respectively. We use coordinates with the Be star as origin in terms of the semi-major axis $a$. Orbital phases are indicated and arrows show the phase of significant events explained in the text.}
\label{fig:orbit}
\end{figure}

The orbital phase at which the type I outbursts of EXO 2030+375 peaks occur has changed dramatically in the past. To investigate the recent evolution of the orbital shift and possible similarities with the past behavior, we used data from MAXI/GSC, Swift/BAT, and RXTE/ASM and determined the peak.\\
The simple Gaussian model used in previous studies \citep{wilson_decade_2002,baykal_recent_2008,wilson_outbursts_2008} did not result in a good fit for many of the outbursts, owing to the better resolution of modern instruments. As known for many BeXRBs \citep[e.g.,][]{2015int..workE..78K}, outbursts can have a variety of shapes. Taking this into account, we used four models, namely the Gaussian model employed in earlier studies (Eq. \ref{eq:gauss}), a Lorentzian model (Eq. \ref{eq:lorentzian}), an asymmetric Gaussian model similar to the one used in \citet{2015int..workE..78K} (Eq. \ref{eq:assymetric}) and some outbursts showing an initial spike were fitted with a double Gaussian model. The peak phase for each outburst was obtained from the model resulting in the best fit,
\begin{equation}
\centering
F_{\textrm{Gaussian}}(x) = K \cdot\exp\left(\frac{-(x-\mu)^2}{2\sigma^2}\right)
\label{eq:gauss}
\end{equation}
\begin{equation}
\centering
F_{\textrm{Lorentzian}}(x) = \frac{K\sigma}{\pi \left(\sigma^2 + \left(x-\mu\right)^2\right)}
\label{eq:lorentzian}
\end{equation}
The parameters $K$, $\mu$, and $\sigma$ specify the scaling factor, the location of the peak, and the standard deviation, respectively:
\begin{equation}
\centering
F_{\textrm{assymetric Gaussian}}(x) = F_{\textrm{max}} 
\begin{cases}
\exp\left(-\frac{\left(x-t_{\textrm{max}}\right)}{2\sigma_{\textrm{rise}}^2}^2\right)\textrm{ for } x < t_{\textrm{max}}\\
\exp\left(-\frac{\left(x-t_{\textrm{max}}\right)}{2\sigma_{\textrm{decline}}^2}^2\right)\textrm{ for } x \geq t_{\textrm{max}}
\end{cases}
\label{eq:assymetric}
\end{equation}
The parameters $F_{\textrm{max}}$, $t_{\textrm{max}}$, $\sigma_{\textrm{rise}}$, and $\sigma_{\textrm{decline}}$ specify the highest value at the peak, the peak time, and the standard deviation of the Gaussian function at the rising and declining phase, respectively.\\
From the fitted peak time, the timing of the peak and hence the orbital phase were determined, as plotted in panel (c) of Fig. \ref{fig:total_plot}. Here, we took $P=46.0205 \pm 0.0002 \textrm{ d}$ as the orbital period and MJD $54044.73 \pm 0.01$ as the origin of periastron \citep{wilson_outbursts_2008}. Data from each monitoring instrument were fitted separately using a range of one orbital period for each outburst (Fig. \ref{fig:precursor}). We verified whether a systematic offset could have been introduced by the use of four different models for the fitting, and found that all models agree within $\pm 1$ d.\\
In panel (c) of Fig. \ref{fig:total_plot} the evolution of the outburst peak can be clearly seen. The outburst peak had a stable phase ~5-6 days after periastron until the OPJ95 (around MJD 50000), when a sudden drop to 4-5 days before periastron occurred, followed by a slow recovery to 3-4 days before periastron. During the giant outburst 2006, a phase shift,OPJ06, of the same amplitude as during the OPJ95 but in the opposite direction (from 3-4 days after periastron to 13-14 days after periastron) was observed \citep{wilson_outbursts_2008}. Afterward, the type I outburst peak slowly changed back to the initial phase 5 of days after periastron.\\ 
The orbital phase of important events is shown in Fig. \ref{fig:orbit}. To calculate the orbit, we used the orbital solution of \citet{wilson_outbursts_2008} and estimated an orbit inclination of $i=48.57$ from the mass function, assuming masses of $M_{\textrm{Be}}=20M_{\sun}$ and $M_{\textrm{NS}}=1.4$ for the companion star and the neutron star, respectively. The calculated orbit and the trajectory of the Lagrangian L1 equilibrium point are shown in the figure. The first calculations of the orbital phase show a stable peak time of the type I outbursts 5.5 days after periastron (orbital phase of 0.12) (A). The majority of type I outbursts in EXO 2030+375 start one week before the peak is reached. For these outbursts, it corresponds to a phase of 0.98 (B). At the time of the OPJ95, the next measured type I outburst suddenly peaked about 5 days before periastron, at a phase of 0.89 (C). From folding the outbursts, we note that most type I outbursts have an approximate duration of one week between the start and the peak time. With this estimate, we can calculate that the outburst after the OPJ95 started approximately at an orbital phase of 0.74. Interestingly, this value is very close to the 2006 giant outburst starting phase of 0.75 (D). We note, however, that the start time of the giant outburst might not have much significance because the outburst lasted for three orbital cycles. We also note that the peak time of the stable type I outbursts (A) and the peak of the outburst after the OPJ95 (C) are $180\degr$ apart.  
\subsection{Outburst precursors}
\begin{figure}
\resizebox{\hsize}{!}{\includegraphics{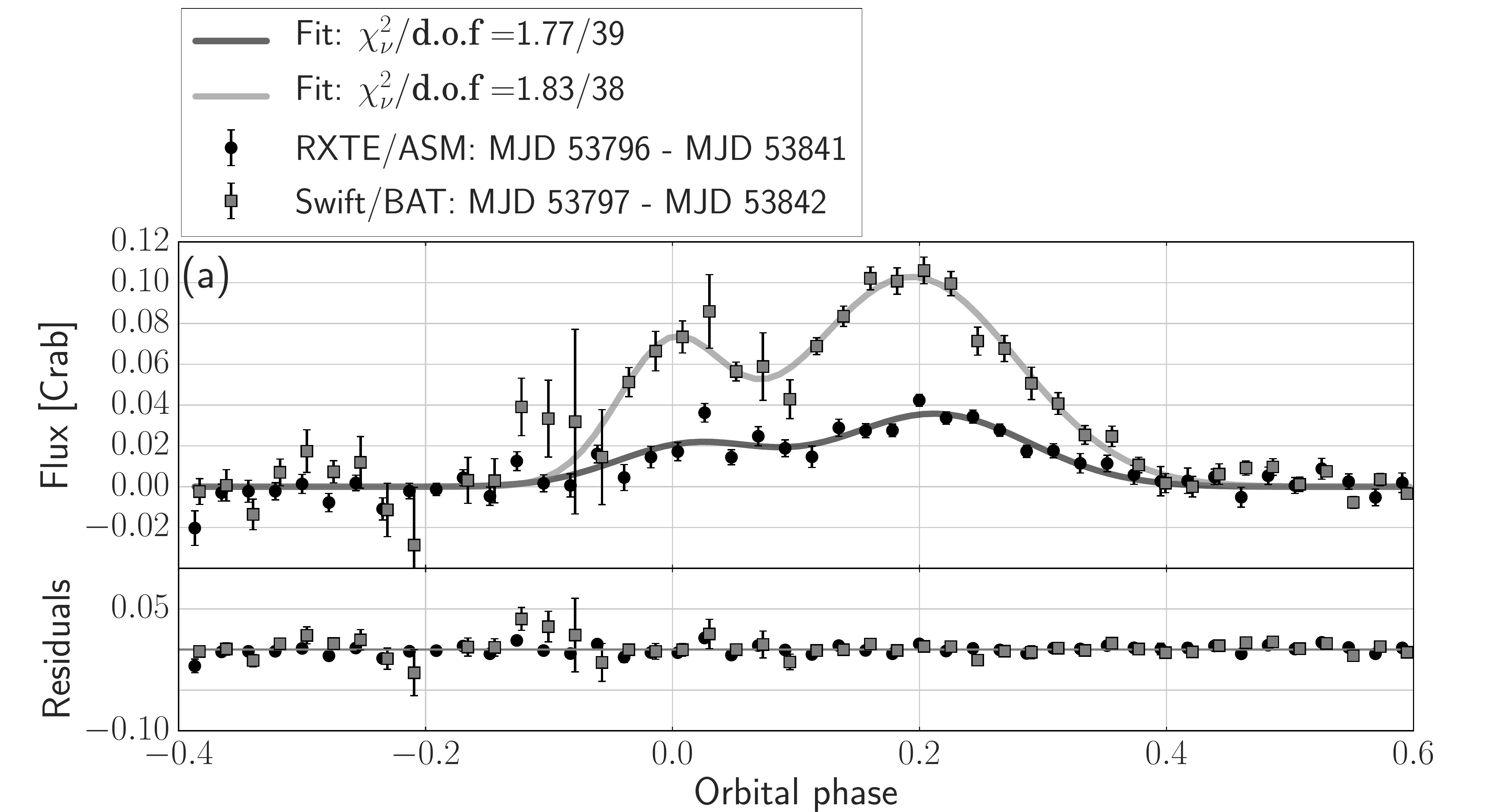}}
	\resizebox{\hsize}{!}{\includegraphics{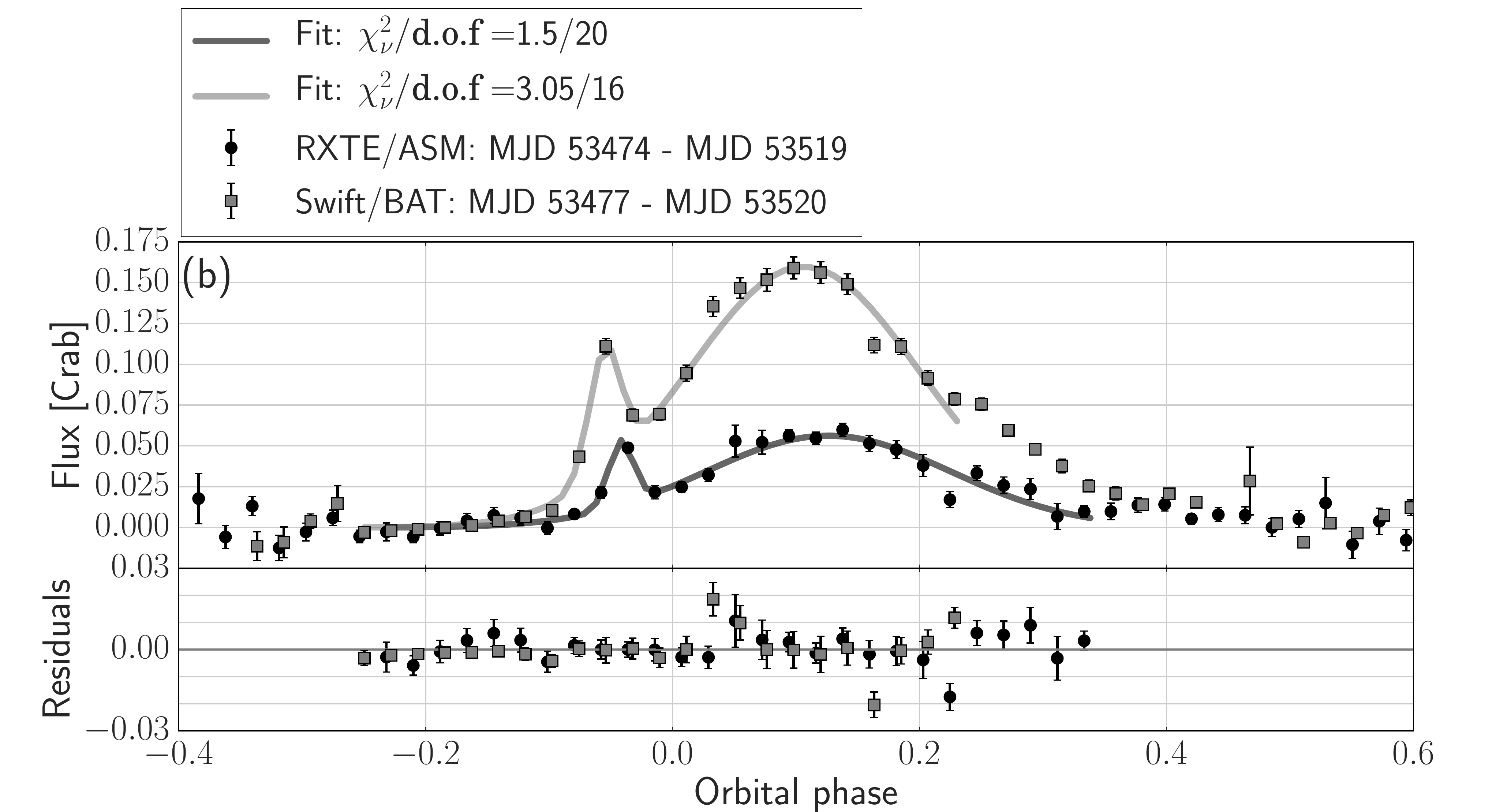}}
\caption{Example of precursor fits with a double Gaussian model for Swift/BAT and RXTE/ASM. (a) Almost double-peaked-like outburst. (b) An initial spike is clearly observed with both instruments.}
\label{fig:precursor}
\end{figure}

For several outbursts before the 2006 giant outburst, an initial spike before the type I outburst peaks was reported in two INTEGRAL observations and three RXTE/PCA observations \citep{camero_arranz_integral_2005}. The authors noted that it appeared to be a recurrent feature.\\
Using MAXI/GSC, Swift/BAT, and RXTE/ASM, we observed that many more outbursts before and after the giant outbursts in 2006 also display this feature. To investigate the time at which a precursor appeared, we fitted these outbursts with the sum of two Gaussian functions (see Eq. \ref{eq:gauss}). The outbursts mainly have two types of precursor shapes: a broad peak and a sharp peak followed by the main outburst. Figure \ref{fig:precursor} shows examples of the two shapes. In panel (a) the case of a first peak with a width almost equal to the next is presented. This outburst might even be classified as a double-peaked outburst. In panel (b) a spike before the regular outburst can be observed. In total, of the 156 outbursts studied, 2 were Fig. \ref{fig:precursor} (a)-like and 17 had a Fig. \ref{fig:precursor} (b)-like precursor.  
As reported in \citet{baykal_recent_2008}, the giant outburst of May 2006 itself (MJD 53900) can also be well approximated by the sum of two Gaussian functions. The resulting precursor peak outburst phase is represented by hollow symbols for each instrument in Fig. \ref{fig:total_plot} panel (c).\\
As shown in Fig. \ref{fig:total_plot} panel (c), the orbital phase of the predecessor outburst can be approximated as being constant around a value consistent with a shift of 4-5 days before the OPJ95 occurred (see also label C in Fig. \ref{fig:orbit}). This might imply that a similar mechanism causes this. It can be argued that the OPJ is a result of a change in the relative strength of the precursor and the main outburst. This argument is contradicted by the fact that no precursors or weak flare in the fading phase was seen before OPJ95 (Fig. \ref{fig:total_plot} c, MJD~49000-50000), when the strongest precursors would be expected.In addition, at the time when the weakest precursors would be expected, around OPJ06 (Fig. \ref{fig:total_plot} c, MJD~53000-54000), the precursors are clearly detected (see, e.g., the penultimate type I outburst before OPJ06, Fig. \ref{fig:precursor} (a)).\\
We did not find the constant interval between the precursor peak and the outburst peak claimed by \citet{camero_arranz_integral_2005}.
\section{Discussion}
\subsection{Periodicity}
\begin{figure*}
\centering
\includegraphics[width=23.5cm, angle=90]{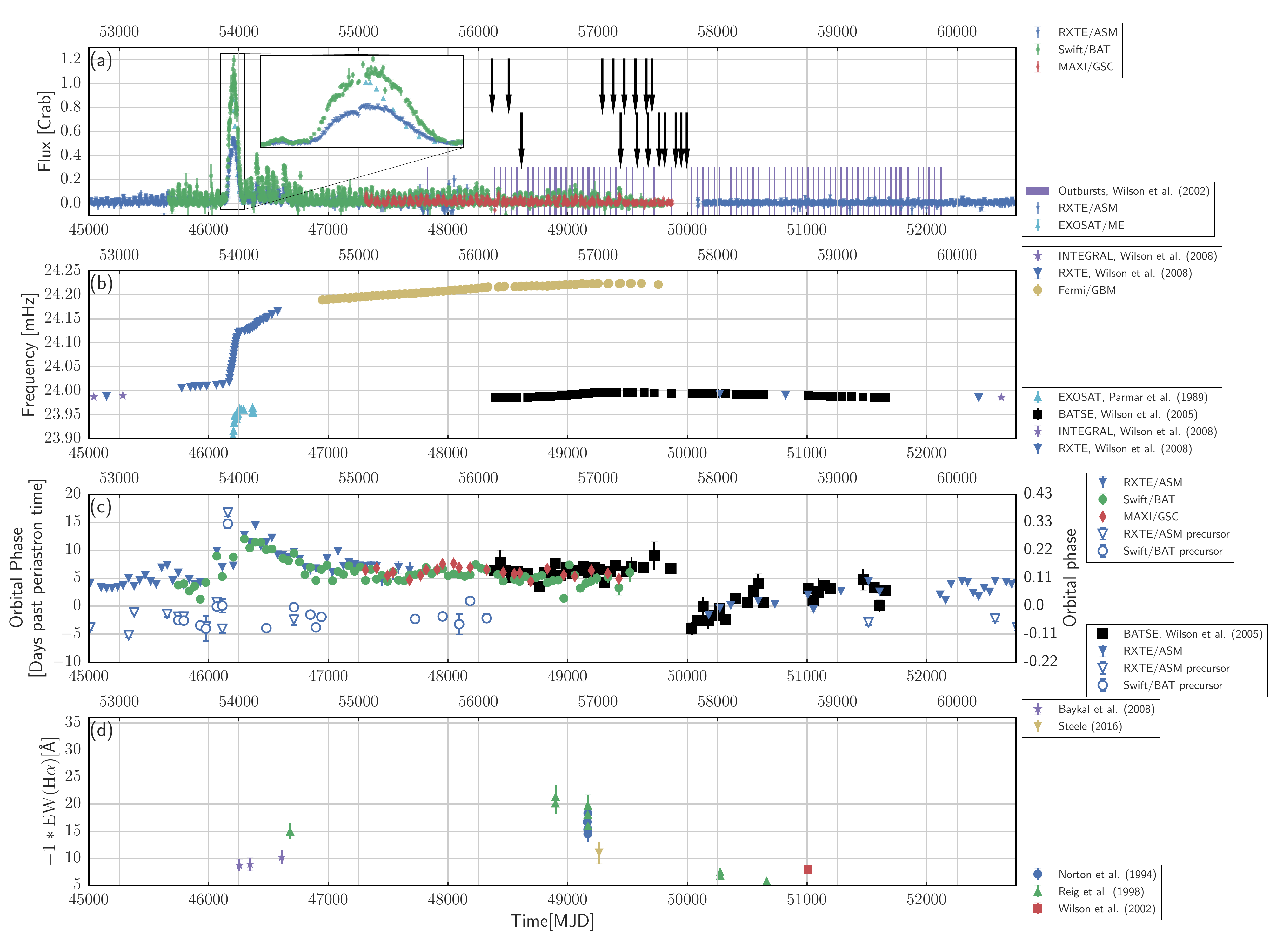}
\caption{Same as Fig. \ref{fig:total_plot}, folded with a period of 7746 d. The times before and after folding are indicated above and below each plot, respectively. On the right side of the plot, legends are shown accordingly. (a) The arrows have been shifted for clarity. The inset shows a zoom of the overlapping giant outbursts.}
\label{fig:folded_plot}
\end{figure*}

\begin{figure}
	\resizebox{\hsize}{!}{\includegraphics{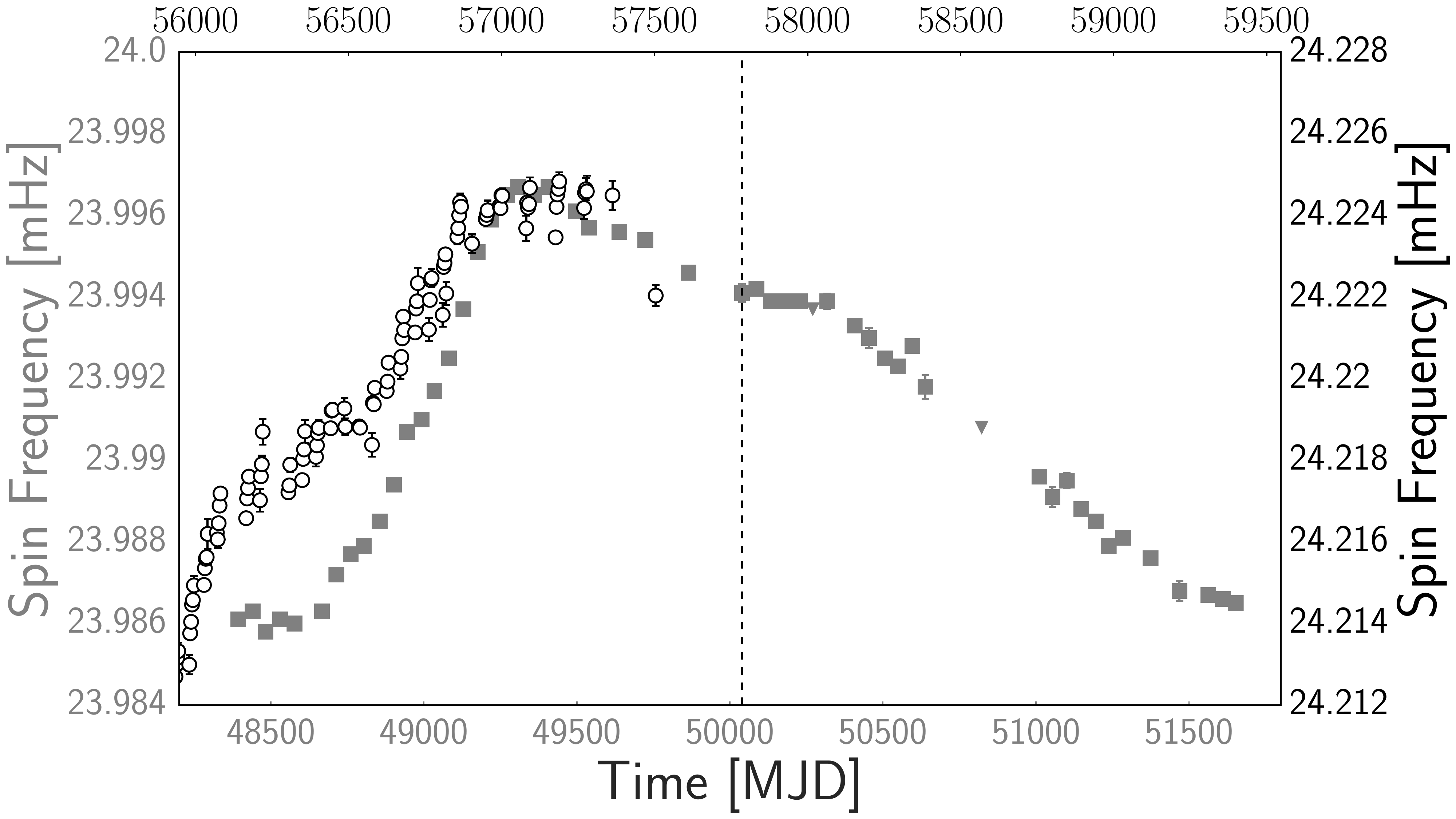}}
	\caption{Spin frequency evolution folded with a period of 7746 d and centered on most recent measurements. The most recent measurements have been shifted for a better comparison, with the same spin frequency scale. Black empty circles, filled gray squares and filled gray inverted triangles correspond to the most recent Fermi/GBM measurements and BATSE and RXTE/ASM measurements from \citet{wilson_discovery_2005}, respectively. The vertical dotted line indicates the time of the OPJ95 and of the predicted orbital phase shift.}
	\label{fig:spin_down}
\end{figure}

The phenomenological similarities between the recent behavior of EXO 2030+375 and the events just before the OPJ95 are striking: missing X-ray outbursts, transition to spin-down and orbital phase of type I outburst peak. They imply analogous underlying physical mechanisms to be present and thus a predictability. Since the events 20 years ago were followed by the OPJ95, it may be natural to expect a similar behavior on the same timescale. We note here that the OPJ95 occurred at a peculiar time, exactly in between the two giant outbursts. This is intriguing and might imply the presence of an underlying periodicity related both to the recently observed events and to the giant outbursts.\\
Assuming the existence of a periodicity, two possible scenarios can be considered:
\begin{enumerate}
\item There is a periodicity in the giant outbursts and orbital phase shifts.\\
\item Only the giant outbursts are periodic, the OPJ95 is a random event.\\
\end{enumerate} 
Both giant outbursts and an orbital phase shift are typically interpreted as related to the interaction between disk and neutron-star . The former results from the accretion of a large amount of matter when the neutron star passes through the disk and the latter from a tilted or precessing disk, leading to capture of matter at a time different from the periastron passage  \citep[e.g.,][]{martin_giant_2014,moritani_precessing_2013,nakajima_precursors_2014}. Therefore, if there is a periodicity, we can expect it to be related to the Be disk state.\\
To estimate the period between giant outbursts, we folded the light curve by matching the declining part of the EXOSAT and RXTE/ASM observations because they have similar energy bands. There is some uncertainty in the derived period because only the decline of the discovery giant outburst was observed. The possibly periodic events and their occurrence times are summarized in Table \ref{tab:events}.\\
\begin{table*}[]
\centering
\caption{Summary of the possibly periodic events and their occurrence time}
\label{tab:events}
\begin{tabular}{ccccc}
\hline\hline
Event                              & 1985 giant outburst & OPJ95 & 2006 giant outburst and OPJ06 & Predicted orbital phase shift \\ \hline
Event start time [MJD] & \textless 46204     & 49870               & 53880               & $57753$                 \\ \hline
Event end time [MJD]   & 46290               & 50040               & 54034               & -                             \\ \hline
\end{tabular}
\end{table*}
From this method, we derived a periodicity of $7746 \textrm{ d}$ between two giant outbursts and consequently $3873 \textrm{ d}$ between the orbital phase shifts and the giant outbursts. Because of the lack of observations for the first giant outburst (at the first detection of the source, the rise was not observed and the peak time is uncertain) and the OPJ95 (no outburst detection before the phase shift for three orbital periods) and as a result of the low number of recurrent events, the uncertainty of the period is strongly  linked to the interval of missing observations and to the assumption used. Assuming half the period to be given by the duration between OPJ95 and OPJ06, the largest uncertainty is estimated to be on the order of the uncertainty on the time at which the OPJ95 occurred, that is, three orbital periods. As noted by \citet{klochkov_giant_2008}, the 1985 giant outburst was certainly brighter than the giant outburst in 2006. Studies of outburst shapes have shown a correlation between brightness and duration of outbursts \citep{2015int..workE..78K}. We can therefore assume that the 1995 giant outburst lasted longer than the 2006 giant outburst. For this reason, OPJ95 most likely occurred almost exactly in between the two giant outbursts. Figure \ref{fig:folded_plot} shows the resulting folded plot (same as Fig. \ref{fig:total_plot}). For clarity, a zoom of the light curve showing the overlapping giant outbursts is inserted. Comparing the previous and recent behavior, we note the following:
\begin{itemize}
\item The spin period behavior (quick turnover after the giant outburst; reaching a plateau after eight years, followed by spin down) and the light curves match very well.\\
\item The recent orbital phase of the type I outburst peak (5 d after periastron) is very similar to the level just before the OPJ95.\\
\item Times of missing outbursts are in good agreement.\\
\item The recent measurement of the H$\alpha$ line is in good agreement with the ones before the OPJ.
\end{itemize} 
With the derived period, we predict the next orbital phase shift to occur at around MJD $57753\textrm{d}$ (December 2016).
Recently, Fermi/GBM data showed strong indications that the source is now starting to spin down. To compare this observation to the frequency measurements from 20 years ago, we folded the spin frequency plot around this time with the derived period of 7746 d. The results can be found in Fig. \ref{fig:spin_down}. The same frequency scale was used, and the recent measurements were shifted at the time in which a plateau was reached for better comparison. The last two measurements have high associated uncertainties because of the current low intensity of the source. Overall, the spin period evolution both now and 20 years ago is very similar, even though the recent spin frequency increase was somewhat faster than the one 20 years ago. The comparable time of spin-down is a strong indication that there is indeed a global periodicity in the behavior of EXO 2030+375 with the calculated period. If confirmed, the latest data point shows that the spin down might be faster this time. 
\subsection{Interpretations}
\subsubsection{Random phenomenon}
We noted that the OPJ95 occurred at a time exactly in between the two giant outbursts. However, because of the low number of observations, we cannot rule out the possibility that this apparent periodicity is due to a random phenomenon. Whether there is a periodicity in the next event and the time at which the next giant outburst occurs can therefore verify this hypothesis.
\subsubsection{Quasi-periodic behavior}
In the early studies of Be X-ray binaries, it was noted that giant outbursts seem to be quasi-periodic. For example, one of the first BeXRB systems, 4U 0115+63, was known for showing a giant outburst every 3-5 years \citep{whitlock_observations_1989,okazaki_natural_2001,boldin_timing_2013}, and the system Cep X-4 has a quasi-period of 4 years \citep{mcbride_cyclotron_2007}. Most explanations of this phenomenon are based on the idea that the Be disk grows until it
reaches a critical size at which a large quantity of matter from the disk can be captured by the neutron star. The duration between giant outbursts is then directly linked to the size and geometry of the Be disk.\\
In the case of EXO 2030+375, the same interpretation can be used for the origin of the giant outbursts, and the OPJ95 can be partly explained by a double Be disk.\\
Several double disks around Be stars have been reported in the past (X Per: \citet{tarasov_double_1995}; $\mu$ Cen:\citet{rivinius_stellar_1998,rivinius_stellar_2001}; 28 CMa and FV CMa \cite{rivinius_evolution_2001}). Of these, X Per is the only BeXRB. \citet{tanaka_dramatic_2007} observed the Be star Pleione using spectroscopy observations and found a double Be disk.\\
The double Be disk consists of an inner disk on the equatorial plane and an outer disk that is highly misaligned and precessing probably due to a companion star.\\
The normal outbursts before the OPJ95 could have been caused at a crossing point of the outer disk and the neutron star orbit. After the Be disk dissipates away even more, there are no outbursts for $\sim 200$ days. This timescale is in agreement with that of other Be disk systems, where changes in size and density of the Be disk were on the order of some hundreds days (e.g., XTE J1946+274 from $-18 \mbox{ \angstrom}$ to $-48 \mbox{ \angstrom}$ in ~100 d \citep{ozbey_detection_2015} and 1A 0535+262 from $1.76 \mbox{ \angstrom}$ to $-10.05 \mbox{ \angstrom}$ in ~430 d \citep{grundstrom_joint_2007}). Then the inner disk reaches the neutron star orbit and, since it rotates on a different plane than the outer one, the crossing point of the neutron star orbit moves and the orbital phase of the type I outburst peak changes.\\
In this scenario, the stop of the mass ejection of the first (outer) disk and start of the second (inner) disk is determined by the activity of the Be star itself. The interval between the first giant outburst and the OPJ95 and that between the OPJ95 and the second giant outburst may not be exactly the same. The outer disk, however, often shows a precession and the orbital phase of the cross section drifts with time. It is not consistent with the observed constant orbital phase of the normal outburst before the OPJ95. Although \citet{wilson_decade_2002} reported the observation of double- and even triple-peaked line profiles, evidence for a double Be disk has not been reported for EXO 2030+375.
\subsubsection{Kozai-Lidov oscillations}
A possible trigger of type II outbursts in BeXRBs, namely Kozai-Lidov oscillations occurring in the Be disk, has recently been proposed \citep{martin_kozai-lidov_2014}. This phenomenon is of particular interest for our study because the oscillations are periodical.\\
Kozai-Lidov oscillations are a physical phenomenon affecting the body with the lowest mass in a three-body scenario under certain conditions, leading to a periodic change of its eccentricity and inclination. They were first described by \citet{kozai_secular_1962} and \citet{lidov_evolution_1962} when the authors solved the problem of the influence of celestial bodies in the solar system (e.g., Jupiter) on the evolution of a test particle orbit (asteroids orbits around the Sun and artificial satellites orbiting planets). When a test particle has a high initial inclination of the particle orbit plane with respect to the binary orbit plane ($39\degr\la i_{p_0}\la 141\degr$) in an initially circular orbit (initial eccentricity $e_{p_0} = 0$), periodic oscillations will appear. As the test particle orbits, its eccentricity and inclination change following the conservation of the the angular momentum component perpendicular to the orbit:
\begin{equation}
\cos{i_p}\sqrt{1 - e_p^2}\approx \textrm{const.}
\end{equation} 
The effect of Kozai-Lidov oscillations on hydrodynamical disks has been simulated for the first time by \citet{martin_kozai-lidov_2014} and was further investigated in \citet{fu_kozailidov_2015,fu_kozailidov_2015-1}. These teams discussed an explicit application to type II outbursts in BeXRBs. By increasing the eccentricity of the circum-stellar disk, the outer edge of the Be disk approaches the neutron star orbit, and then the neutron star can capture matter directly from the Be disk, even when not at periastron \citep{martin_giant_2014}. Using hydrodynamical disk simulations, the authors found that Kozai-Lidov oscillations can occur, although they become damped. Assuming a rigid disk with radius $R$ and a surface density $\Sigma$ following a power law $\Sigma \propto R^{-p}$, the timescale of the oscillations $\tau_{\textrm{KL}}$ can be approximated as
\begin{equation}
\frac{\tau_{\textrm{KL}}}{P_{\textrm{orb}}}\approx \frac{\left(4-p\right)}{\left(\frac{5}{2}-p\right)}\left(\frac{a}{R_{\textrm{out}}}\right)^{\frac{3}{2}}\sqrt{\frac{M_{\textrm{Be}}}{M_{\textrm{NS}}}\left(\frac{M_{\textrm{Be}}}{M_{\textrm{NS}}} + 1\right)}\textrm{ ,}
\end{equation}
with $M_{\textrm{Be}}$ and $M_{\textrm{NS}}$ the mass of the Be star and the neutron star, respectively; $P_{\textrm{orb}}$ the binary orbital period, $a$ the binary separation in a circular orbit, and $R_{\textrm{out}}$ the initial disk outer radius. However, this estimate does not take into account the dependence on the inclination angle. As was discussed in \citet{fu_kozailidov_2015}, the timescale of Kozai-Lidov oscillations becomes longer, the start time slightly later, and the amplitude smaller with a decreasing inclination angle. The authors show that the equation is valid to about a factor 2.\\ 
Within this accuracy, the outer disk radius can be approximated by the binary separation at periastron $D_{\textrm{peri}}$ and the orbital separation by the semi-major axis $a_{\textrm{sm}}$. Following simple ellipse geometry, we can express both quantities as a function of the eccentricity $e$ of the neutron star orbit:
\begin{equation}
\frac{a}{R_{\textrm{out}}}\approx\frac{a_{\textrm{sm}}}{D_{\textrm{peri}}} = \left(1-e\right)^{-1}.
\end{equation}
For a realistic estimate of the timescale of Kozai-Lidov oscillations in the Be disk of EXO 2030+375, we assume reasonable values for the unknown parameters: $p=1.5$ \citep{martin_kozai-lidov_2014}, $M_{\textrm{Be}}=20 M_\sun$, and $M_{\textrm{NS}}=1.4 M_\sun$ \citep{okazaki_natural_2001}. Using the orbital parameters described in \cite{wilson_outbursts_2008}, we obtain $\tau_{\textrm{KL}} \approx 83~P_{\textrm{orb}} \approx 3820 \textrm{ d}$. Considering the uncertainties of the parameters, this estimate is similar to the period we derived between an orbital phase shift and a giant outburst.
\subsubsection{Kozai-Lidov oscillations in EXO 2030+375}
As pictured in Fig. 3 of \citet{martin_kozai-lidov_2014}, Kozai-Lidov oscillations in hydrodynamical disks lead to a periodic exchange of the Be disk eccentricity for its inclination. Both parameters experience a damped oscillation between two extrema, with the eccentricity maximum occurring at the inclination minimum and vice versa.\\
From the previous calculation and the known uncertainty of the timescale estimation, we can consider two interpretations of a periodic change in EXO 2030+375 that are linked:
\begin{enumerate}[(a)]
\item The giant outbursts and the OPJ95 occur when the eccentricity of the disk is at its maximum. In this case, the time between a giant outburst and an orbital phase shift corresponds to the oscillation timescale, as our calculation suggests.\\
\item Only the giant outbursts occur when the eccentricity of the disk is at its maximum and the OPJ95 corresponds to the eccentricity minimum, that is, the inclination angle maximum. Therefore, the Kozai-Lidov oscillation timescale is of ten years, which is within the accuracy of the calculated value of between 3820d (10.5 y) and 7640d (20.9 y).\\
\end{enumerate}
In interpretation (a), the disk eccentricity becomes higher around MJD 46000 and extends far enough (beyond the truncation radius) for the neutron star to accrete a large amount of matter, leading to a fast spin-up and a giant X-ray outburst. The same scenario can explain the giant outburst around MJD 53900. In this case, we know from observations that the giant outburst started around the orbital phase 0.75, the same phase as the start of the OPJ95 (see Fig. \ref{fig:orbit} and explanations in Sect. \ref{section:peakTiming}). We can identify this phase as the start phase of the crossing region between the disk and the neutron star orbit. The differences between the OPJ95 and the giant outbursts must therefore originate in another effect that affects both the optical and the X-ray flux, such as disk loss or precession.\\
We consider the possibility of a disk loss after the previous giant outburst. The decrease of H-alpha equivalent width, near-infrared brightness, and the fading in X-rays suggest that the disk size and/or density gradually decreased around MJD~49000. However, because of the Kozai-Lidov effect, the disk eccentricity increases while the density of the disk decreases, which enables the neutron star to pass nearby the outer disk at a given phase. This leads to the observation of an orbital phase shift instead of a giant outburst. This is supported by the similar starting orbital phase of the giant outburst and of the OPJ95 and the similar equivalent width of the H$\alpha$ line measurements. From infrared data \citep{wilson_outbursts_2008}, it seems that the color did not change, contrary to what we expect for a disk-loss scenario.\\
In interpretation (b), the giant outbursts occurs for the same reason as in the previous case: the enhanced eccentricity of the Be disk enables the accretion of a large quantity of matter by the neutron star and hence a giant X-ray outburst apart from periastron. However, in this case, the OPJ95 is due to the changes in the Be circum-stellar disk inclination, which reaches a maximum at the time of the orbital phase shift. As a consequence of the inclination change, the projected area of the Be disk becomes smaller, leading to the observed drop in the equivalent width of the H$\alpha$ line and infrared brightness. At the same time, since the disk eccentricity becomes lower and the inclination increases, the outer region of the disk is farther away from the neutron star, leading to a drop in the accretion rate and hence in the X-ray flux. In this model, the OPJ95 could be caused by the inclination change. With this interpretation, it remains unclear why the timescale of changes in the orbital phase of the type I outbursts' peak after the OPJ95 and the timescale of the H$\alpha$ variability is much shorter than the timescale of the Kozai-Lidov oscillations.
\subsubsection{Predictions}
If there is an underlying periodicity due to Kozai-Lidov oscillations, then we expect certain phenomena to occur in the next years, depending on the model used:
\begin{itemize}
\item In interpretation (a), we expect a giant outburst, accompanied by an orbital phase jump, to occur every 10 or 20 years. If the disk is large enough, then we would expect a giant outburst to occur approximately in December 2016. If not, then we may observe an orbital phase shift. Additionally, this model implies that a giant outburst should be observed approximately in August/September 2027.In the double-disk scenario, the same prediction is made with less accuracy.\\
\item Interpretation (b) predicts a jump in the orbital phase of the type I outburst peak in around December 2016 and a giant outburst starting approximately in August/September 2027.\\
\end{itemize}

\subsection{Note added after submission}
An orbital phase shift occurred in EXO 2030+375 \citep{2016ATel.9263....1L}. The outburst reached a peak on 2016 July 20 at an orbital phase of 0.015. Compared to the almost constant peak phase of 0.13 in the past 7 years (see Fig. \ref{fig:total_plot} (c)), this represents a shift of ~5 days in the same direction as for the OPJ95. This event, which we had predicted, supports our suggestions of a 20-year periodicity, even though it did not occur at the predicted time (end of December 2016).\\
It is interesting to note that the amplitude of this orbital phase shift has only half the value of OPJ95 and OPJ06. This suggests that we are only seeing the beginning of an orbital phase shift that might further increase during the next type I outbursts. This might be due to the better resolution of the current X-ray monitors.\\
Observations of the behavior of the source during the type I outbursts following this event in the X-ray, optical, and infrared range will clarify whether interpretation (a) or (b) is more plausible (future work).
\section{Conclusions}
The main findings of our study can be summarized as follows:
   \begin{itemize}
      \item Phenomenologically, the recent unusual behavior of EXO 2030+375 is very similar to the events before MJD 50000: (1) a drop in the X-ray flux with some outbursts becoming undetected; (2) a transition from slow spin-up to almost constant spin, then spin-down; (3) a strengthening of the H$\alpha$ line equivalent width.  
      \item From all these similarities, a ~3800-day or ~7700-day periodicity of events can be derived, which can be related to the occurrence of type II outbursts and to orbital phase changes. If this periodicity exists, we predict an orbital phase shift to occur around December 2016. At the same time, we might observe a giant outburst.
      \item We have shown that such a periodicity could be plausibly explained by Kozai-Lidov oscillations in the Be disk.
      \item Recent X-ray observations seem to support the existence of a 20-year periodicity.
   \end{itemize}
Infrared and optical observations are needed to verify the Be disk state and to confirm whether a recurrence exists.
\begin{acknowledgements}
This research has made use of the MAXI data provided by RIKEN, JAXA and the MAXI team.; Swift/BAT transient monitor results
provided by the Swift/BAT team and data from the Fermi/GBM pulsar project. The authors thank Iain Steele for his help. E.L. is grateful for the warm hospitality of the MAXI team and thanks the IPA program in RIKEN for the support.
\end{acknowledgements}

%
%
\bibliographystyle{aa} 
\bibliography{EXO.bib}
\end{document}